\documentclass[useAMS,usenatbib]{mn2e}

\title[Natural orbits]
{Natural orbit approximations in single power-law potentials}
\author[C. Struck] 
{Curtis Struck \thanks{E-mail: curt@iastate.edu} \\
Department of Physics and Astronomy, Iowa State University, Ames, IA, 50014 USA}
\usepackage{amssymb}
\usepackage{amsmath}
\usepackage{graphicx}
\usepackage{multirow} 
\def\aap{{ A\&A}}

\def\aj{{AJ}}

\def\apj{{ApJ}}

\def\mnras{{MNRAS}}

\begin{document}
\date{\today}

\pagerange{\pageref{firstpage}--\pageref{lastpage}} \pubyear{0000}

\maketitle

\label{firstpage}
\begin{abstract}

In a previous paper, I demonstrated the accuracy of simple, precessing, power ellipse (p-ellipse)  approximations to orbits of low-to-moderate eccentricity in power-law potentials. Here I explore several extensions of these approximations to improve accuracy, especially for nearly radial orbits. 1) It is found that moderately improved orbital fits can be achieved with higher order perturbation expansions (in eccentricity), with the addition of `harmonic' terms to the solution. 2) Alternately, a matching of the extreme radial excursions of an orbit can be imposed, and a more accurate estimate of the eccentricity parameter is obtained. However, the error in the precession frequency is usually increased. 3) A correction function of small magnitude corrects the frequency problem. With this correction, even first order approximations yield excellent fits at quite high eccentricity over a range of potential indices that includes flat and falling rotation curve cases. 4) Adding a first harmonic term to fit the breadth of the orbital loops, and determining the fundamental and harmonic coefficients by matching to three orbital positions further improves the fit. With a couple of additional small corrections one obtains excellent fits to orbits with radial ranges of more than a thousand for some potentials. 

These simple corrections to the basic p-ellipse are basically in the form of several successive approximations, and can provide high accuracy. They suggest new results including that the apsidal precession rate scales approximately as $log(1-e)$ at very high eccentricities $e$. New insights are also provided on the occurrence of periodic orbits in various potentials, especially at high eccentricity. 
 
\end{abstract}


\section{Introduction}

Celestial mechanics in the Solar system is grounded on the Kepler ellipse solution for the point-mass potential, and perturbations to it. Components of galaxies present a number of other spherically-symmetric potentials, some of which can be approximated by single or multiple power-law forms. Unfortunately, no exact orbital solutions, like the Kepler ellipse, are known for any but a handful of these cases. 

This paper will focus on a series of approximations to fit individual orbits accurately, even nearly radial orbits, in a range of spherically symmetric potentials. Although orbit solutions are an ancient problem, there have been a number of recent publications on sophisticated methods to approximate them in the literature; some of these were noted in \citet [henceforth Paper I] {st06}, and \citet{va12}. Most of this attention has focused on orbits in asymmetric potentials, e.g., relevant to barred galaxies, (but see \citet{re10} on periodic orbits in symmetric potentials.) The logarithmic potential, which in its symmetric form describes flat rotation curve discs, has been the object of special attention. \citet{be07} and \citet{pu08} have used Lie transform methods and normal form approximations to find periodic orbits. \citet{co90} and \citet{va12} expanded on the \citet{pr82} orbit approximation. \citet{tt97} developed a symplectic map technique to study general orbital dynamics in the logarithmic and other potentials. 

Another area of focus in these and other papers (e.g., \citealt{va05}) is determining the rate of apsidal precession in various potentials. The classical theorem of \citet{ne87} gives this for spherical power-law potentials in the limit of near circular orbits. \citet{va05} and \citet{va12} numerically integrate expressions they derive for the eccentricity dependence of less circular orbits, especially in the logarithmic potential. The latter paper emphasizes the utility of the Lambert W and polylogarithmic functions. Most of these techniques involve series solutions, below I will develop a compact, accurate, analytic approximation for the precession function. 

\citet [also see \citealt{ly08}] {ly10} used action-angle methods to derive an analytic expression for the orbits in any central potential, with ``better than 1 per cent accuracy for the angle between pericentre and apocentre,'' though less accurate for highly eccentric orbits in some cases. (Also see \citealt{ue14}.) \citet{wi14} build on this work, and derive approximations with very small errors in the actions, and which become exact in the circular and radial limits. They apply their results to stellar streams; \citet{sa14}also apply an action space approximation to galactic tidal streams. That application will not be considered in this paper, but the very accurate approximations given for orbits with very large radial could prove very useful there.

In Paper I, I presented the basic properties of a very simple family of approximations to orbits in power-law potentials, especially potentials relevant to galaxy components, which are extended here. The functions considered were powers of the azimuthal part of the usual polar equation for a Kepler ellipse, and were called p-ellipses. In addition to doing a good job of modeling the radial excursions of orbits up to moderate eccentricities, these p-ellipses are surprisingly good at approximating the rates of apsidal precession over a significant range of power-law index in the potentials. The methods explored in this paper have many points of contact with the literature, especially with those of \citet{ly10}. The methods detailed in the latter apply to a wider range of potentials, but the approximations described below appear to do at least as well in fitting an extensive range of orbits in power-law potentials, and are based on the simple p-ellipse functions rather than approximations to an integral for the azimuthal angle. 

An aside on terminology - the term p-ellipse is conveniently brief, if not very descriptive. It is tempting to use the term ``power ellipse,'' since this term is more like the naming of other generalizations of the simple ellipse function, including the superellipses (Lam\'e curves) and the ``superformula'' (or Gielis) curves (\citealt{gi03}). However, ``power ellipse'' is not any more descriptive, and is potentially confusing. The p-ellipses are generally different than either of the other ellipse generalizations, though they can produce many similar oval or star-like forms. Some particular p-ellipses with added harmonics (as described below) are special cases of the superformula. The p-ellipses are also related to, but not the same as epicycloids and spirograph curves.

To return to the topic of approximation, perturbation expansions of the p-ellipse solution in eccentricity provide relations between the several parameters of the solution. The results of Paper I were largely obtained with a first-order perturbation approximation. It was briefly noted in Paper I that by freely adjusting the three parameters (see below) of a p-ellipse, one could do much better in fitting a wide range of orbits, including those with high eccentricities. This is one reason p-ellipses can be considered as ``natural'' orbit solutions in power-law potentials. However, these results leave us with a gap between having a systematic procedure for fitting orbits over a range of potential index at low eccentricity, and a trial-and-error procedure otherwise. As a specific example, the first-order approximation predicts that the apsidal precession frequency (relative to the circular frequency) depends only on the power-law index of the potential, yet it is clear from numerical models that it also depends on the radial range or eccentricity for high values of these quantities (see \citealt{va05}). Of course, there are a wide range of general, multi-parameter functional fitting techniques available to fit individual orbits, but they do not give the valuable scaling information, such as the dependence of precession frequency on eccentricity. 

A straightforward method is to extend perturbation approximations to higher order, and in the following we will explore second and third order approximations, though to avoid being over-determined, these require additional terms in the solution. Series approximations to orbits have no such difficulty, since they have infinite terms. The present case is more like epicyclic approximations, where successive improvements require added epicycles. Although exactly what to add, and how to add it is not as clear. In addition, the effect on accuracy of imposing different initial or boundary range conditions in the parameter relations is also considered below. The default is specifying the position and velocity of the orbiting body at an initial time. However, other choices are possible, and it turns out that they are usually more productive. 

In the following section (and the Appendix) I derive or summarize the equations of motion, the p-ellipse solution and perturbation approximations to various orders. In Section 3 this toolkit is used to compare the results of various approximations in a range of cases. Section 4 considers the special cases of orbits with extreme radial ranges, and modifications of the approximations to deal with them. In Sections 3 and 4 a rather long series of approximations is explored. Successful prescriptions are discovered, but some unproductive or even unstable alternatives are also noted.  The results are summarized and applications outlined in the final section.

\section{Equations and approximations in single component potentials}

\subsection{Basic equations}

In this section I present the basic equations used in the remainder of the paper.  Newton{'}s equation of motion for a massless particle in a fixed, spherically symmetric power-law potential can be written, 

\begin{equation}
\label{eq1}
\frac{d^2r}{dt^2} =  g_r + \frac{v^2_\phi}{r},\\
\end{equation}
\begin{equation*}
g_r = -\frac{GM}{r^{2\delta + 1}},\ \ \delta = \frac{1}{2};\\
\end{equation*}
\begin{equation*}
g_r = -\frac{GM_\epsilon \epsilon^{2\delta - 1}}{r^{2\delta + 1}},
\ \ \delta \neq \frac{1}{2};
\end{equation*}

\noindent
where $r$ is the particle{'}s radius in the orbital plane, $v_\phi$ its azimuthal velocity, and $\delta$ is the power specifying the potential. In the first case ($\delta = 1/2$) $M$ is the total mass, in the second one there exists an arbitrary scale radius here designated $\epsilon$, and $M_\epsilon$ is the mass contained within it.  This notation is generally consistent with Paper I, though there the scale radius was identified with a softening radius. In this paper softening of the power-law potential is not considered. As in Paper I, we replace time, $t$, as the independent variable with azimuth $\phi$. After the substitution equation \eqref{eq1} becomes, 

\begin{equation}
\label{eq2}
uu'' =  {c_\delta}u^{2\delta} - u^2,
\end{equation}
\begin{equation*}
u = r^{-1},\ c_{\delta} = \frac{GM}{h^2},\  \delta = 1/2;
\end{equation*}
\begin{equation*}
u = \epsilon r^{-1},\ c_{\delta} = \frac{GM_{\epsilon}\epsilon}{h^2},\  \delta \neq 1/2;
\end{equation*}

\noindent where the double prime notation indicates the second derivative with respect to $\phi$, and with specific angular momentum $h$.

The basic p-ellipse solution of Paper I is, 

\begin{equation}
\label{eq3}
u = \frac{1}{p} \left[ 1 +
e \cos \left( m{\phi} \right) 
\right]^{\frac{1}{2} + \delta},
\end{equation}

\noindent
where this function is recognized as a simple ellipse with the part in square brackets taken to the power $\frac{1}{2} + \delta$. For a fixed potential this solution has three parameters: $e$, the eccentricity, $p$, the semi-latus rectum, and $m$, which is the ratio of precession and orbital frequencies. Sometimes it will be referred to simply as the frequency. When $\delta = \frac{1}{2}$ and $m = 1$, then equation \eqref{eq3} gives the Kepler ellipse. 

A three-dimensional parameter space is not huge, especially with initial conditions, conservation conditions or extremal values of $u$ to constrain the parameter values. Perturbation theory can provide additional relations needed to solve for the parameter values, or alternately, a basis for comparison to other approximations.

\subsection{Perturbation approximations}

In Paper I the eccentricity was assumed to be small, and approximations were derived by expanding equation \eqref{eq3} in powers of $e$. There the focus was on the first-order approximation, while here we extend consideration to the second and third-order approximations in order to study more eccentric orbits. 

\subsubsection{First-order approximation}

We recall from Paper I that the first-order approximation gave equations for the semi-latus rectum of an orbit as a function of the orbit constant $c_\delta$ (equation \eqref{eq2}), and of the orbital frequency $m$, in terms of the power index $\delta$. That is, to first-order, 

\begin{equation}
\label{eq4}
c_\delta = \left(\frac{1}{p_1}\right)^{2\left(1-\delta\right)},
\ \ \ \ 
m_1 = \sqrt{2\left(1-\delta\right)},
\end{equation}

\noindent Note that $m_1$ is also the ratio of the classical epicyclic frequency to the orbital frequency. A formula for the eccentricity in terms of $p$, and the specific angular momentum was also given in Paper I. 

\subsubsection{Second-order approximation}

A second order expansion of the p-ellipse equation \eqref{eq3}, after substitution into the equation of motion gives three equations for the coefficients at constant, first, and second-order in  $ecos(m\phi)$. However, because the p-ellipse is so simple, with only two parameters ($m$ and $p$) to solve for, this solution is over-constrained. The solution must be generalized. A variety of possible generalizations have been explored. For example, in Paper I an epicycle of the same frequency was added to the power ellipse. This attempt yielded no general improvements in the approximation. 

To match orbital shape and the precession frequency, another approach is to consider harmonics, which would modify the shape of the orbit within a precession period. Furthermore, if the harmonics are of higher order in $e$, then they would couple to higher order terms from the expansion of the simple p-ellipse. I.e., we should consider terms of the form, $e^ncos(nm\phi)$, where $n$ is a whole number. Such harmonic terms are related to powers of the fundamental, e.g., $cos^2(m\phi) = (1 + cos(2m\phi))/2$, and it turns out to be more convenient to use these powers in perturbative expansions of the p-ellipse. Specifically, the we choose the following generalization of equation \eqref{eq3} as the basis of a second-order approximation,

\begin{equation}
\label{eq5}
u = \frac{1}{p} \left[ 1 +
e \cos \left( m{\phi} \right)
+ f_2 e^2 cos^2\left( m{\phi} \right) 
\right]^{\frac{1}{2} + \delta},
\end{equation}

\noindent which includes the additional coefficient, $f_2$. Expanding this function to second-order we obtain,

\begin{multline}
\label{eq6}
up =  1 +
\left({\frac{1}{2} + \delta}\right) e \cos \left( m{\phi} \right)\\
+ \frac{1}{2} \left({\frac{1}{2} + \delta}\right) 
\left(2f_2 - \frac{1}{2} + \delta\right)
e^2 cos^2\left( m{\phi} \right) .
\end{multline}

\noindent After substitution of this expression and its second derivative into the equation of motion (equation \eqref{eq2}), and some algebra, we get the following coefficient equations. First,

\begin{equation}
\label{eq7}
\frac{c_\delta}{p_2^{2\left(\delta-1\right)}} =
\frac{1 - \left({\frac{1}{2} + \delta}\right)  \left(\frac{1}{2}  - \delta - 2f_2\right) e^2}
{1 +  \left(1-2\delta\right)  \left({\frac{1}{2} + \delta}\right)
\left(\frac{1}{2}  - \delta - 2f_2\right) e^2},
\end{equation}

\noindent for $p_2$, the second-order semi-latus rectum and then,

\begin{multline}
\label{eq8}
m_2^2 = 1 + \left(1-2\delta\right) \frac{c_\delta}{p_2^{2\left(\delta-1\right)}} = 1\\
+  \left(1-2\delta\right)
\left[ \frac{1 - \left({\frac{1}{2} + \delta}\right)  \left(\frac{1}{2}  - \delta - 2f_2\right) e^2}
{1 +  \left(1-2\delta\right)  \left({\frac{1}{2} + \delta}\right)
\left(\frac{1}{2}  - \delta - 2f_2\right) e^2} \right], 
\end{multline}

\noindent for the second-order frequency. Combining all three coefficient equations yields the following equation for $f_2$ in terms of $\delta$ and $e$ alone,

\begin{multline}
\label{eq9}
f_2 = \frac{1}{4} - \frac{1}{2}\delta\\ 
-\left[ \frac{ \left({\frac{1}{2} - \delta}\right) \left({1 + \delta - 2 \delta^2}\right)}
{2 \left[3 - 3\delta + \left({\frac{1}{4} - \delta^2}\right)
\left({1 + \delta - 2 \delta^2}\right) e^2 \right]} \right],
\end{multline}

\noindent which completes the second-order approximation.

\subsubsection{Third-order approximation}

The procedure for deriving the third-order approximation is the same as that of the previous section, if more complex. To begin, we add another harmonic term to equation \eqref{eq5} to obtain,

\begin{multline}
\label{eq10}
u = \frac{1}{p} \times\\ \left[ 1 +
e \cos \left( m{\phi} \right)
+ h_2 e^2 cos^2\left( m{\phi} \right)
+ h_3 e^3 cos^3\left( m{\phi} \right) 
\right]^{\frac{1}{2} + \delta},
\end{multline}

\noindent Here we use the letter $h$ for the coefficients, instead of $f$, to distinguish the different cases, i.e.,  $f_2$ does not equal $h_2$. The third-order expansion in $ecos(m\phi)$ is,

\begin{multline}
\label{eq11}
up =  1 +
\left( \frac{1}{2} + \delta\right) e \cos \left( m{\phi} \right)\\
+ \frac{1}{2} \left( \frac{1}{2} + \delta \right) 
\left( 2h_2 - \frac{1}{2} + \delta\right)
e^2 cos^2 \left( m{\phi} \right)\\
 + \frac{1}{6} \left( \frac{1}{2} + \delta \right) 
\left(\frac{-3}{2} + \delta \right) 
\left(2h_2 - \frac{1}{2} + \delta\right) \times\\
\left(6h_3 + 4\delta h_2 \right)
e^3 cos^3\left( m{\phi} \right).
\end{multline}

\noindent As before this expression and its second derivative (with respect to $\phi$) are substituted into the equation of motion, and in this case the result is a set of four coefficient equations. These four equations can in turn be solved for the four variables: $m$, $p$, $h_2$, and $h_3$.  These various equations, which complete the third-order approximation, are given in the appendix.

To anticipate the results below, we will find that these harmonic approximations are useful aids in the search for accurate orbit approximations. However, their direct application turns out not to yield rapidly convergent solutions in many cases; another approach is more effective.

\begin{figure}
\centerline{
\includegraphics[bb=150 20 450 750, scale=0.65]{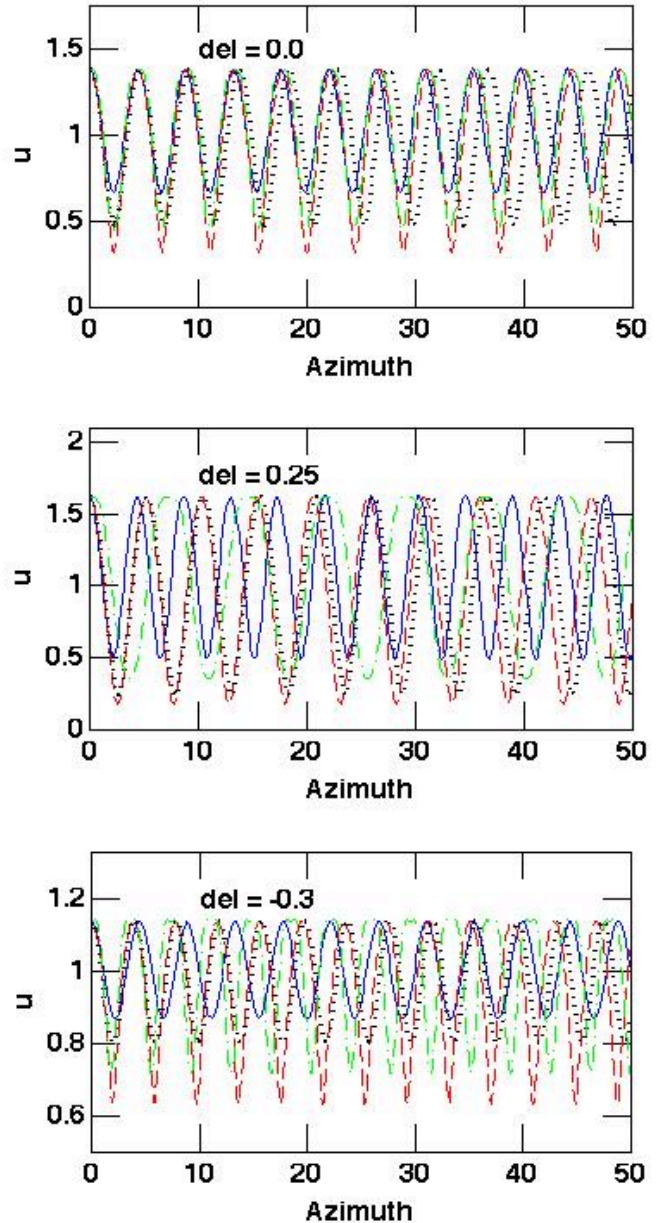}}
\caption{The three panels show the fit of perturbative approximations in three different potentials: the $\delta = 0$ logarithmic potential on the top, the $\delta = 0.25$ falling rotation curve potential in the middle, and the $\delta = -0.3$, rising rotation curve potential on the bottom. The blue curve gives a numerical integration of the orbit for the given initial radius and radial velocity (i.e., 0.0). The red, dashed curve gives the first-order, p-ellipse approximation, while the black dotted curve gives a second-order approximation with one harmonic term added to the p-ellipse, and the green dot-dash curve gives a third-order approximation with two added harmonics. See text for details.  }
\end{figure}

\begin{figure}
\centerline{
\includegraphics[bb=150 50 450 750, scale=0.65]{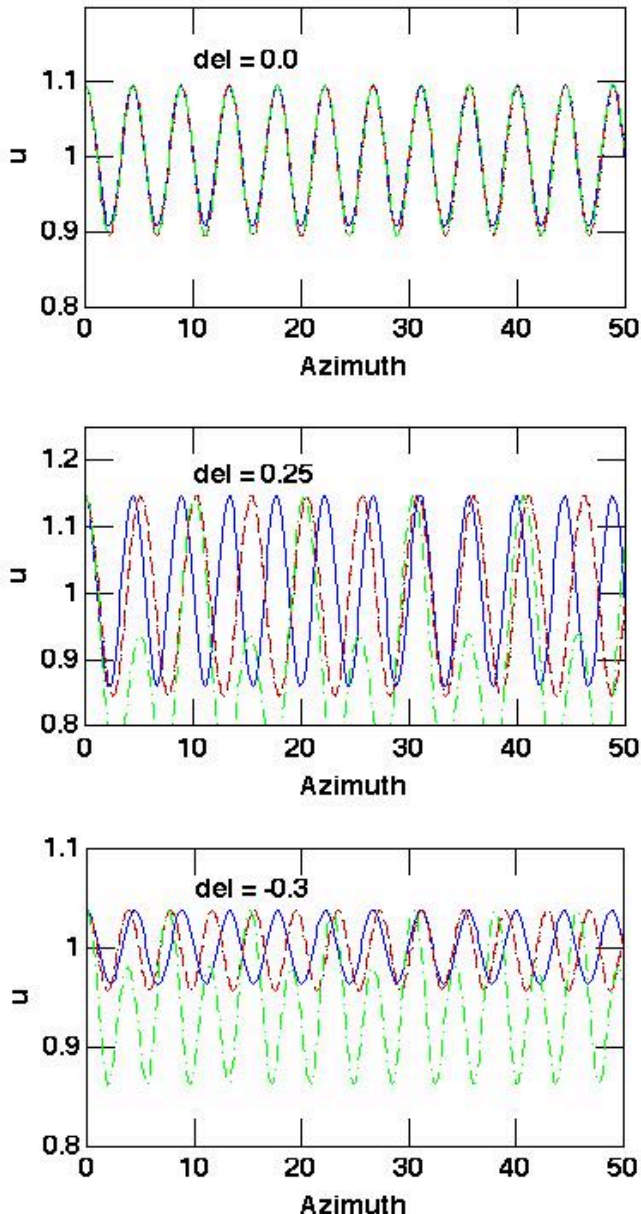}}
\caption{Same as Fig. 1, except with orbits of smaller radial range, showing convergence in the first-order approximations, and lack of convergence in the higher order approximations in some potentials.}
\end{figure}

\section{Comparisons between approximate and numerical orbits}

In this section we consider accuracy of several different extensions of the first-order, p-ellipse approximation, specifically, using added harmonic terms with and without perturbation expansions. 

\subsection{Perturbation approximations with simple initial conditions}

In Paper I the p-ellipse approximations and numerical orbits were compared within the context of several specializations. First, it was assumed that the initial conditions were such that the orbiting particle was located at the highest value of $u(\phi = 0)$ with zero radial velocity (periapse of $r$). These particular initial values only effect the fit in the trivial manner that it will be best at high values of $u$ and worst at low values. As we will see below, the fact that we choose initial conditions of position and velocity at one point, rather than matching the radial range of the orbit is more consequential. 

A second specialization in Paper I was that the value of the semi-latus rectum was generally chosen to be $p = 1$. This was viewed as a convenience, but actually raises a more subtle issue. The p-ellipse orbits always have a symmetry in $u$ around the value $u = 1/p$, i.e., $u'' = 0$ when $u$ has this value ($m\phi = \pi/2, 3\pi/2$, etc.). This is not generally true of the numerically integrated orbits, except when the force constant is $c_\delta = 1$. Evidently, $c_\delta$ does not scale with $p$ exactly as predicted by the p-ellipse approximation of equation \eqref{eq3}, but has a more complicated dependence on $p$. However, this problem can be circumvented by using a specific selection of units, where $c_\delta = 1$. This can be achieved by adjusting the units of the semi-latus rectum, so that it will be the unit of length to first order. A particular orbit approximation can be scaled to a given orbit size after the other parameters are determined.

With these conventions, we can then compare the approximations of the previous section to the numerical orbits. Fig. 1 gives a first example. It shows a comparison between the three levels of approximation of the previous section and the numerical orbit in the cases with three different potentials, and orbits extending over about a factor of a few in radius. The initial variable values were the same in all approximations shown in each panel.

The top panel shows the radial ($u$) variation over about a dozen orbital periods in a representative flat rotation curve ($\delta = 0.0$) case. Clearly, the first order approximation (red dashed curve) goes to much lower values of $u$ (larger radii) than the numerical solution (blue solid curve).  The second-order approximation (black dotted curve) partially corrects this defect, but drifts more rapidly in phase relative to the numerical solution than the first order approximation. The third order approximation largely corrects this phase effect, but does not improve the amplitude error. If these trends continued with additional harmonic terms, then it appears that one would have to go to fifth-order or higher in the perturbation expansion to obtain a very accurate approximation for an orbit with this amount of radial variation. 

The middle panel shows a similar case in the declining rotation curve potential with index ($\delta = 0.25$)  half way between the Kepler value ($\delta = 0.5$) and the flat rotation curve cases. Here, the results are qualitatively similar to the previous case. However, the second-order solution does not correct the amplitude error much, and the phase drift of the third order solution is not reduced as much as in the previous case. Of course, this harmonic addition and perturbation expansion procedure would also not be the best way to fit an ellipse in the Kepler case. Indeed, only the first-order approximation (equation \eqref{eq4}) gets the frequency $m$ correct in that case. We will return to this point below.

The lower panel shows a rising rotation curve case, with ($\delta = -0.3$). In such cases, the radial variations are generally quite constrained for given initial conditions, and the first order approximation is quite poor at capturing this. The second-order approximation does much better. The third order approximation does not do better in correcting either amplitude or phase. In the rising rotation curve cases the phase drift is in the opposite sense of that in declining rotation curves.  The poor performance (and relatively large expansion harmonic  coefficient) of the third order approximation, suggests that this perturbation approach may not converge in such cases, or if it does converge, it may not do so monotonically. 

One aspect of convergence is partially addressed in Fig. 2, which shows similar plots to Fig. 1, but for orbits with much smaller radial ranges or eccentricities. The top panel shows that all of them converge nicely to the numerical orbit at low eccentricity in the flat rotation curve case. The convergence is not as good (in both amplitude and phase) in the declining rotation curve case shown in the middle panel. In fact, the amplitude appears not to be converging, judging by the third-order approximation. 

This is also true of the rising rotation curve case shown in the lower panel. In this panel the amplitude estimate of the first and second approximation is quite good, though the rate of phase drift is high.

In sum, the added harmonic plus perturbation expansion is useful for modeling orbits with given starting points in potentials near the $\delta = 0.0$ case. In these potentials the third order approximation clearly provides a better fit, especially over long time periods,. For potentials far from this case, it appears that the third order corrections are too large (not perturbative), and the approximation sequence may diverge in most cases. Overall, this procedure seems to yield rather limited improvements. This is not entirely surprising given that the convergence of the perturbation expansion is based on small values of the eccentricity. Because the coefficients in the expansion are not always small, evidently this procedure cannot be extended to relatively high values of eccentricity (or even moderate values in cases far from $\delta = 0.0$).

\subsection{Fitting the radial range, but using perturbation equations for the precession rate}

The perturbation procedure of the previous section is not how an ellipse is fitted to an orbit in the Kepler problem. In that case, the orbit (but not its orientation and the particle position on it) is specified by two parameters, which can be either the eccentricity and the semimajor axis, the specific energy and angular momentum, or the inner and outer radii. These elementary facts are worth recalling because they can be generalized for p-ellipses. 

Specifically, we obtain an equation for the eccentricity parameter as the ratio of two instances of equation \eqref{eq3} evaluated at the inner and outer radii, where $m\phi = 0, \pi$. This equation can be solved to obtain,

\begin{equation}
\label{eq12}
e = \frac{u_{+}^{\frac{2}{1+2\delta}} - u_{-}^{\frac{2}{1+2\delta}}}
{u_{+}^{\frac{2}{1+2\delta}} + u_{-}^{\frac{2}{1+2\delta}}},
\end{equation}

\noindent where $u_{+}$ is the inner radius (largest $u$ value), and $u_-$ is the outer radius (also see \citealt{ly10}, equation( 7)). Evaluating equation \eqref{eq3} again at $u_{+}$ we obtain an expression for $p$, 

\begin{equation}
\label{eq13}
p = \frac{\left( 1+e \right)^{\frac{1}{2}+\delta}}
{u_+}.
\end{equation}

\noindent Analogous, but implicit expressions, can be obtained for the solutions with one or two added harmonics. 

Of course, p-ellipses depend on three parameters, rather than two, so the orbit is not completely specified by the radial range or the energy and angular momentum. We need to take the precession into account, that is, find a value for $m$. One way to do that is to use the formulae from perturbation equations, that is, either equation \eqref{eq4}, \eqref{eq8}, or \eqref{eq26} for the first, second, or third order approximations (with harmonic terms in the solution), respectively. 

Fig. 3 shows the results of this exercise, for the same three potentials and in the same format as Fig. 1. Clearly, the fit is generally better than Fig. 1, if simply because the radial amplitude is forced to fit. The phase drift is not generally better, however, though it is different. This is mostly because we have retained the overlarge radial range of the first-order approximation of Fig. 1, via the perturbation equations, so the eccentricities are too large, yielding incorrect values of $m(e)$ from the second and third-order formulae. The main point is that as before the approximate solutions drift within a few orbital periods, except in the case of the first-order approximation to the rising rotation curve case (lower panel). Even that result does not obtain for all rising rotation curve potentials. 

\begin{figure}
\centerline{
\includegraphics[scale=0.65]{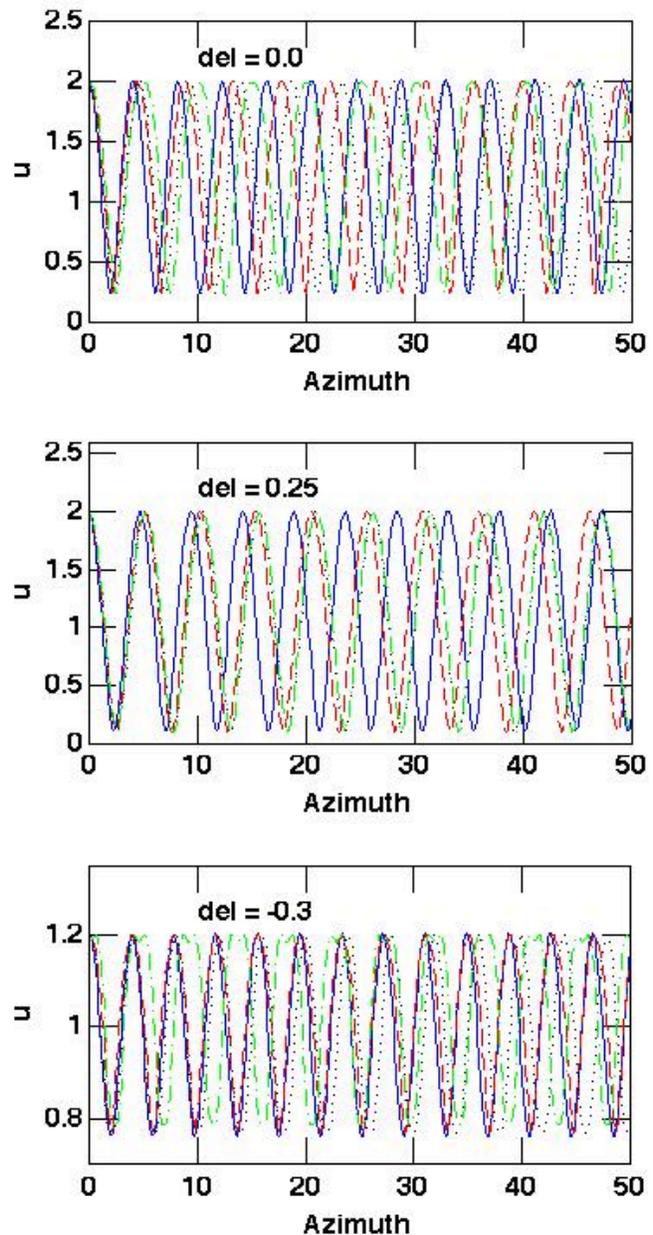}}
\caption{Same as Figs. 1 and 2, but in the case where p-ellipse parameters $e$ and $p$ are determined by the radial range rather than the initial values, though the precession frequency $m$ is determined by the coefficient equations of the perturbation expansion. See the text for details.}
\end{figure}

\begin{figure}
\centerline{
\includegraphics[scale=0.45]{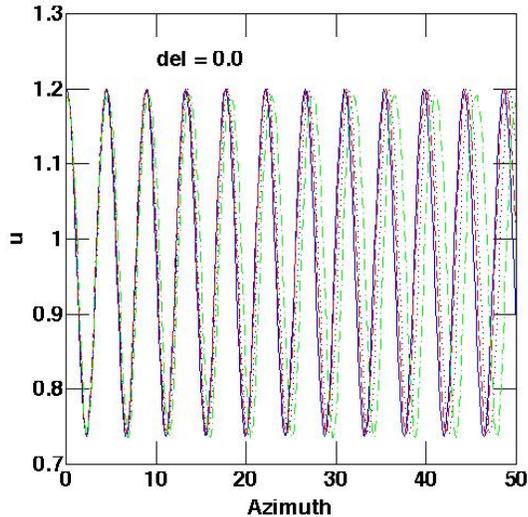}}
\caption{Shows an example using the same approximations as used to produce Fig. 3, but with an orbit of small radial range. All three approximations for $m$, of different orders, converge in this limit, in this potential, but the third-order does not converge as well as the others. }
\end{figure}

Fig. 4 is analogous to Fig. 2, that is, showing orbits of small radial amplitude in this case. Of course, the amplitudes of the higher order approximations do not diverge in this case, but there is still some phase drift at low values of $e$. 

\subsection{Fitting the radial range and correcting the first-order precession rate} 

There is another way to estimate the third orbital parameter, $m$. That is simply to find a  correction function, e.g., from numerically integrated orbits, to the first-order estimate of equation \eqref{eq4}. This might appear to be the same as individual orbit fitting, which we abjured above. However, Figs. 1-4 show that the first-order estimate is quite accurate, so we can expect that the correction is generally modest, and hope that it is a smooth and slowly varying function across orbits with varying $e$ and $\delta$ (as suggested by the work of  \citealt{va05}, and \citealt{va12}). If so, we can approximate the correction with a simple analytic fit, and thus, obtain a  good analytic approximation. 

Fig. 5 shows sample orbits in four different potentials, with the blue solid curves showing the numerically integrated orbits, and the dashed red curves the fitted p-ellipse approximations. The dotted curves show p-ellipse with an added harmonic discussed below. All of these are fairly radial orbits, and the fits are generally very good, and with the adjusted value of $m$, it remains good over an arbitrarily long time.

\begin{figure}
\centerline{
\includegraphics[scale=0.42]{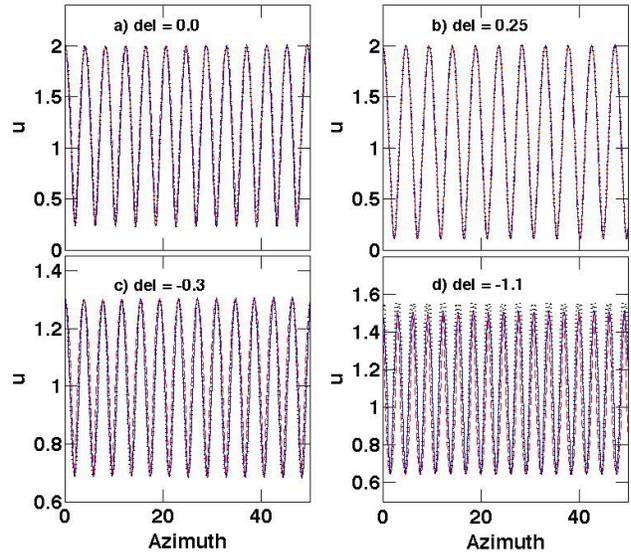}}
\caption{Same as the previous $u-\phi$  figures with the numerical integration again shown as solid blue curves. However, in this case the first approximation (red, dashed curves) is derived by fitting the radial range, and the using the precession rate correction of equations \eqref{eq14}. The second approximation (black dotted curve) adds a harmonic term, fit to a third point on the orbit, to better capture the orbit?s shape. Panels a-d show a sample orbit in potentials with indices: $\delta = 0.0, 0.25, -0.30, and -1.1$, respectively.}
\end{figure} 

\begin{figure}
\centerline{
\includegraphics[scale=0.42]{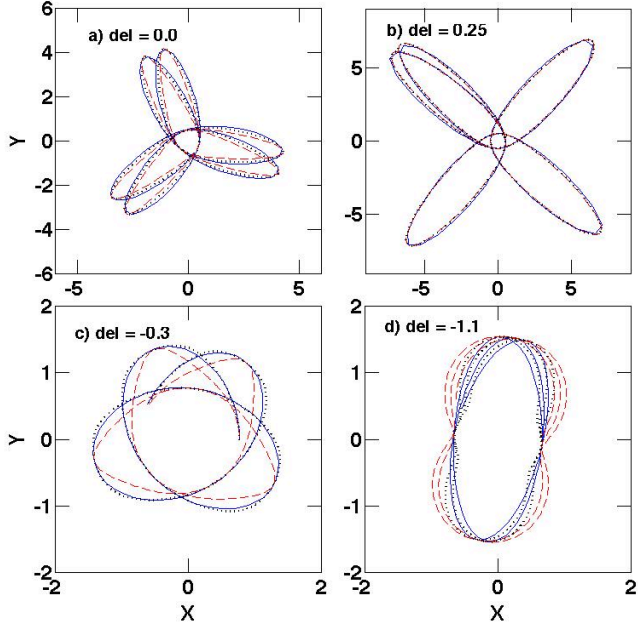}}
\caption{Same as Fig. 5, but showing the motion in the $x-y$ orbital plane.}
\end{figure}

A few features of these individual cases are worth noting. First, Fig. 5a demonstrates that in the logarithmic potential this simple approximation can be quite accurate even for orbits spanning a large radial range (nearly a factor of 10 in the case shown). Secondly, for potentials closer to the Kepler case, like that shown in Fig. 5b, the accuracy is even better. Thirdly, the approximation is much weaker in the case of small negative values of $\delta$, like that shown in Fig. 5c. The greater deviation of the dashed power ellipse curve from the numerical curve is clear, though the dotted harmonic curve is still a good fit. The radial range is not so large in this case, and the approximation gets worse rapidly as that range is increased. Fig. 5d shows a nearly solid-body rotation curve case where the fit is also not highly accurate, though still reasonable. In this case the coefficient of the added harmonic becomes large enough to add extra `wiggles' at large $u$ values, which also expands the radial range of that approximation. 

The fits of Fig. 5 were achieved initially by trial-and-error adjustments of $m$. In each of these potentials many more such fits were undertaken for orbits over a large range of eccentricity. It was found that the modest corrections to $m$ are generally smooth functions of the variable $x = log(1-e)$, and these functions could, in turn, be well approximated by least-squares fits to fairly low-order polynomials. Specifically, if $m_{corr}$ is the corrected value, and $m_o$ is the initial value of $m$ (from equation \eqref{eq4}), then we find, for the cases illustrated in Fig. 5,

\begin{multline*}
\frac{m_{corr}}{m_o} = 1.0013 - 0.00439x + 0.0520x^2
+ 0.0169x^3\\ + 0.00180x^4, \delta = 0.0;
\end{multline*}
\begin{multline*}
\frac{m_{corr}}{m_o} = 0.9997 - 0.0181x + 0.0476x^2
+ 0.0124x^3, \delta = 0.25;
\end{multline*}
\begin{multline*}
\frac{m_{corr}}{m_o} = 1.000 - 0.00179x + 0.0077x^2
+ 0.0015x^3, \delta = -0.30,
\end{multline*}
\begin{equation}
\label{eq14}
with\ \ x = log_{10}\left( 1-e \right).
\end{equation}

\noindent  in all cases. These formulae are quite accurate up to orbits with radial ranges slightly greater than those shown in Fig. 5. They are a bit cumbersome, but can be simplified with the loss of a little accuracy. First of all, we note that the corrections are negligible for the Kepler and solid-body cases ($\delta = 1/2, -1$, respectively), and small for other rising rotation curve potentials between the flat rotational curve and solid-body cases. Secondly, note that the expressions in the first two expressions of equations \eqref{eq14} are quite similar. Thus, with linear fits to the potentials with $\delta = 0.25, 0.0$, and $-0.30$, we obtain,

\begin{equation*}
\frac{m_{corr}}{m_o} \simeq 1.000,\ \   \delta = 0.5;
\end{equation*}
\begin{equation*}
\frac{m_{corr}}{m_o} \simeq 0.99 - 0.065x,\ \   \delta = 0.0 - 0.25;
\end{equation*}
\begin{equation*}
\frac{m_{corr}}{m_o} \simeq 0.998 - 0.010x,\ \   \delta = -0.30;
\end{equation*}
\begin{equation}
\label{eq15}
\frac{m_{corr}}{m_o} \simeq 1.000,\ \   \delta = -1.0.
\end{equation}

\noindent Linear interpolation should be reasonably accurate for $\delta$ values between those listed. 

\subsection{Harmonics again to improve the fit accuracy}

The radial range and precession frequency fitting procedure of the last section works very well over a significant range of potential index and eccentricity, and much better than the perturbation expansions of Section 3.1 and 3.2 (Figs. 1, 2). Is there a systematic procedure for further improvements? The fact that the dashed curves in Fig. 5 tend to deviate most from the numerical solution at orbital phases of around $\pi/2$ and $3\pi/2$ suggests that adding a first harmonic term as in equation \eqref{eq5} might improve the fit. 

Recall that the harmonic terms were introduced in Section 2 to provide extra terms needed for the perturbation expansion approach. The coefficient equations of that expansion provided formulae for the coefficients of harmonic terms. In the present context we must match another point on the true orbit to determine the harmonic coefficient. We will choose the value at $m\phi = \pi/2$, which in the present examples will be obtained from the numerically integrated orbit (or alternately from the conservation conditions).

It is convenient to change the notation of this harmonic power ellipse slightly from that of Section 2, i.e., 

\begin{equation}
\label{eq16}
u = \frac{1}{p} \left[ 1 +
e_1 \cos \left( m{\phi} \right)
+ e_2 cos\left( 2m{\phi} \right) 
\right]^{\frac{1}{2} + \delta},
\end{equation}

\noindent In adopting this form we are incorporating the harmonic in terms of $cos(2m\phi)$ rather than a $cos^2(m\phi)$ term, and are no longer assuming that the harmonic coefficient is of the order of $e^2$ (or in this case $e_1^2$). Then the parameters $p, e_1$, and $e_2$ are determined by matching the solution at the maximum and minimum values of $u$ ($u_+$ and $u_-$ again), and $u(m\phi = \pi/2)$, or $u_{\pi/2}$, i.e., 

\begin{multline}
\label{eq17}
pu_+ = \left[1+e_1+e_2 \right]^{\frac{1}{2}+\delta},\ \ or,\ \ 
\left( pu_+ \right)^{\frac{2}{1+2\delta}} = 1+e_1+e_2 ,\\
pu_- = \left[1-e_1+e_2 \right]^{\frac{1}{2}+\delta},\ \ or,\ \ 
\left( pu_- \right)^{\frac{2}{1+2\delta}} = 1-e_1+e_2 ,\\
pu_{\pi/2} = \left[1-e_2 \right]^{\frac{1}{2}+\delta},\ \ or,\ \ 
\left( pu_{\pi/2} \right)^{\frac{2}{1+2\delta}} = 1-e_2 .
\end{multline}

\noindent The latter forms can be divided (the first by the second, second by the third), to obtain two equations for $e_1$ and $e_2$ ($p$ cancels) in terms of ratios of the given values of $u$. The solutions are,

\begin{equation}
\label{eq18}
e_1 = \frac{1}{2}\left( e_+ + e_- \right),\ \ 
e_2 = \frac{1}{2}\left( e_+ - e_- \right),
\end{equation}

\noindent with,

\begin{equation*}
e_- = \frac{1 + \frac{1}{2} \left[ \left(\frac{u_+}{u_{\pi/2}} \right)
^{\frac{2}{1+2\delta}}
- 3 \left(\frac{u_-}{u_{\pi/2}} \right)
^{\frac{2}{1+2\delta}} \right]}
{1 + \frac{1}{2} \left[ \left(\frac{u_+}{u_{\pi/2}} \right)
^{\frac{2}{1+2\delta}}
+ \left(\frac{u_-}{u_{\pi/2}} \right)
^{\frac{2}{1+2\delta}} \right]},
\end{equation*}
\begin{equation}
\label{eq19}
e_+ = \left( 1-e_- \right) \left( \frac{u_+}{u_-} \right)
^{\frac{2}{1+2\delta}} - 1.
\end{equation}

\noindent Then $p$ is determined by substituting these results into any of the equations \eqref{eq17}. The frequency ratio $m$ is determined as in the previous subsection.

This was the procedure used to produce the black dotted curves in Fig. 5. In Figs. 5a and 5b the dotted curves fit the numerical integration so well they are hard to see. They are a little more visible in Fig. 5c. As already noted, in Fig. 5d the harmonic coefficient is large enough that it introduces a wiggle in the fitted curve, but otherwise the fit is reasonable.

Fig. 6 provides another view with plots of the same orbits and fits as in Fig. 5, but in the orbital plane. In Fig. 5 the radial differences between the dashed curve power ellipse and the numerical integration are visually minimized. Not so in Fig. 6, but the fits are still generally quite good, at least in Figs. 6a-c. Note that in the numerical integration the Matlab routine ODE45 was allowed to use its default choice of azimuthal interval, resulting in a choppy form at large radii.  Note that the curve in Fig. 6b is nearly periodic. In potentials with aperiodic orbits at low eccentricity, as $m_{corr}(e)$ changes with increasing radial ranges it will pass through rational values, yielding closed orbits (see equations \eqref{eq14}). Its value in the case shown in Fig. 6b is close to 4/3. (Fig. 6a is also not far from such a resonance.)

In Fig. 6d we see the extra wiggles of the dotted curve partially counter the tendency of the power ellipse to be `squeezed out,' away from the numerical curve at azimuths farthest from the ones determining the fit. This success inspires the idea that adding more harmonics might lead to an even better fit in difficult solid-body type cases. Some exploratory work was attempted in that direction, but it revealed a fundamental problem. If the harmonic coefficients are not very small, then the sum of 1.0, the fundamental and the harmonic terms in generalizations of equation \eqref{eq16} may equal zero or a negative number, yielding a break-down of the approximation. This appears to be a common result of adding even one more harmonic. 

It may be that there is a natural combination of harmonics that avoids this difficulty. Since equations \eqref{eq17} and their generalizations are linear, this could be investigated systematically with many harmonics using a numerical Cramers' method to solve for the harmonic coefficients. This is beyond the scope of this work.

To summarize this section, several methods of extending the p-ellipse approximation to better fit more radial orbits have been explored. Expansions in the eccentricity parameter from single point initial conditions were marginally successful (Section 3.1). However, fitting approximation parameters to the extreme radii yields a better result (Section 3.2), a point also made in \citet{ly10}. This likely relates to the fact that it is equivalent to matching the conserved quantities of the orbit. Orbit fits are further improved with the precession frequency correction factor of equations \eqref{eq14} and \eqref{eq15} (Section 3.3).  The best approximation over a range of power indices comes from adding a harmonic term to the power ellipse (Section 3.4). With this harmonic term, we effectively use another orbital point constraint to better fit the orbit's shape. 

\section{Extreme fits to nearly radial orbits}

The best approximations of the previous section yielded good fits to quite radial orbits over most of the relevant range of power index. Remarkably, comparable fits can be obtained for orbits that are much closer to being purely radial. The difference from the above discussion is that because any inaccuracy in the approximation parameters can give a noticeable difference in the radial range, these parameters must be specified quite accurately.

Fig. 7 shows an example in the logarithmic potential of an orbit that ranges over a factor of more than 130 in radius. The computed eccentricity of the power ellipse approximation is $e = 0.99991$. A fitted value of $m = 1.2345$ is used in the figure, while the value predicted from equation \eqref{eq14} is $m = 1.2333$. This difference of one digit in the fourth place results in a small, but noticeable phase drift after about half a dozen orbits. The point is that a little extra care is needed to accurately fit p-ellipse orbits that are nearly radial, but the adjustments from the formulae of the previous section may not be very large. 

\begin{figure}
\centerline{
\includegraphics[scale=0.44]{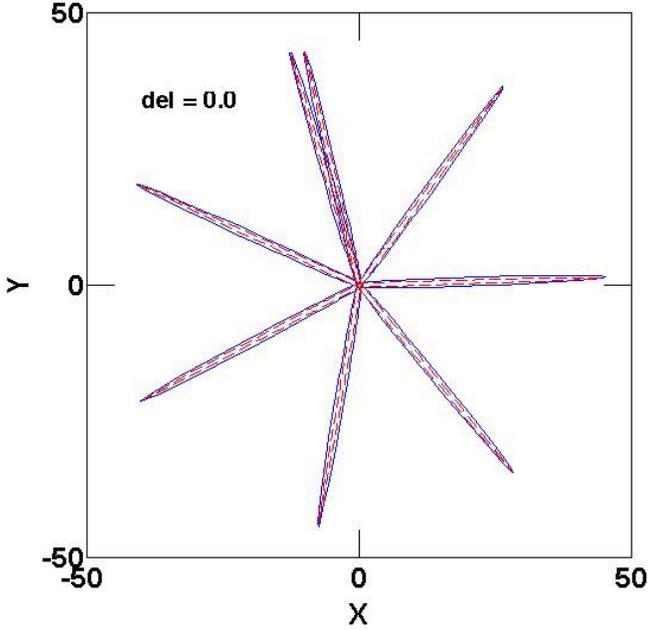}}
\caption{A very radial orbit (factor of 130 in radial range), with the numerical integration shown by the solid curve. The red, dashed curve shows a p-ellipse approximation fitted according to the prescription of Section 3.4, with additional fine-tuning of the precession frequency as described in the text. }
\end{figure}

This is good, but it is possible to do even better. Fig. 8 shows a fit to the orbit of Fig. 7 with a p-ellipse plus one harmonic as in equation \eqref{eq16}. In this figure the numerical integration is a blue solid curve, while the approximation is a black solid curve, and except at a few points it is very difficult to distinguish the two curves over the eight orbits shown. Although this approximation is of the form of equation \eqref{eq16}, if one uses the procedure of Section 3.4 to solve for the parameters, one obtains a curve that extends to more than twice the radius shown. And yet, in a certain sense that is a good fit. That is, the approximation adheres closely to the numerical orbit over most of its course. Unfortunately, the `wiggles' that appear at high values of the parameter $e_2$ (e.g., Fig. 5d), are very large in this case, and corrupt the solution. 

However, the wiggles can be tamed by smaller values of $e_2$, and since the solution is very sensitive to precise parameter values for these nearly radial orbits, it is likely that the procedure above simply does not yield sufficiently accurate estimates of the parameters. For example, the decision to use the value of the orbit at $m\phi = \pi/2\ (u_{\pi/2})$ may not be optimal. Further experimentation seemed warranted, and that yielded the following discoveries.  

The simple p-ellipse solution of Fig. 7 is enveloped by the true orbit. The harmonic term expands the approximate solution to better fit the true orbit. A larger value of $e_2$ yields more expansion, roughly. However, a large value of that parameter can lead to a breakdown in the solution in the manner discussed in Section 3.4, unless we require that $e_2 \le e_1 - 1$. Actually, we generally want near equality in this relation. Through this relation, the value of $e_1 - 1$ determines the width of the orbital lobe of the approximate orbit. The exact value of $e_2 - e_1$ determines its radial extent. So too does the value of $p$, but this parameter can be calculated from the initial value of $u_+$, as before. (Other values of $u$ are not used since we are now fitting the $e$ parameters rather than deriving them.)

For example, to produce the approximation of Fig. 8, the values used are $e_1 = 1.24$, and $e_2 = 0.240133$. In these parameters the digits `24' determine the width of the curve, and the digits `133' determine its radial extent (in conjunction with $p$). 

\begin{figure}
\centerline{
\includegraphics[scale=0.44]{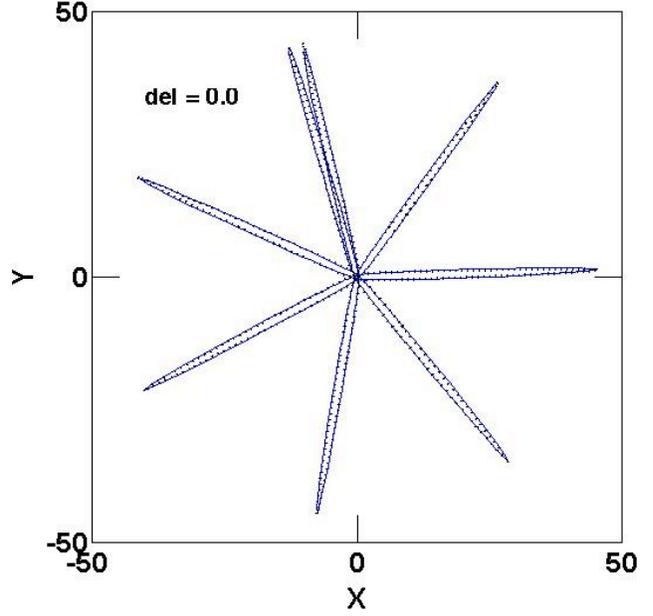}}
\caption{Same as Fig. 7, but in this case the p-ellipse approximation has an added harmonic, and further coefficient corrections as described in Section 4. The approximation is drawn as a solid blue curve, which is virtually indistinguishable from the dotted black numerical curve here. }
\end{figure}

The upshot of this exercise is that it points the way to very simple, yet very accurate approximations for nearly radial orbits. The example of Fig. 8 demonstrates this. A number of other cases were computed, in the logarithmic potential, in order to derive a correction function to improve the values of $e_1$ obtained from equations \eqref{eq18} and \eqref{eq19} above. The result, to be added to the value derived from those equations is,

\begin{multline}
\label{eq20}
\Delta e_1 = 0.562 - 0.792y + 0.304y^2
- 0.061y^3 + 0.0050y^4,\\ with\  \ y = log_{10} \left( \frac{u_+}{u_-} \right).
\end{multline}

\noindent The correction is quite accurate for orbits with radial ranges of $10 \le u_+/u_- \le 10^4$. 

A functional fit was also derived for the very small quantity $1 + e_2 - e_1$, which determines the fit to the radial range, again for the logarithmic potential. As a function of the same variable $y$ as in equation \eqref{eq20} it is,

\begin{equation}
\label{eq21}
log_{10} \left(1+e_2-e_1 \right) = 0.410 - 2.024y.
\end{equation}

\noindent Given the coefficient of $y$, this is essentially a quadratic relation, valid over the same range as the previous. (And given the excellence of the fit, probably over a larger range as well.) These corrections can also be derived for other potentials, or in the case of small, positive power-law indices, interpolated between the Kepler problem (with $e_2 = 0.0$) and the above relations. Judging from Fig. 8 and other cases not shown, the resulting corrected p-ellipse plus harmonic forms fit the conserved quantities very well.

\section{Summary and Outlook}

With appropriate choices of the parameter values, p-ellipses can accurately, and persistently fit orbits in power-law potentials over a wide range of eccentricity and power-law index. This accuracy, their simplicity (equation \eqref{eq3}) or \eqref{eq16}), and their close relation to the ellipse solution to the Kepler problem, earn them the designation ``natural'' approximations. However, these functions have more advantages to recommend them over, e.g., circular epicycles and some series approximations. Foremost among these is that there exists a kind of separable and optimal procedure for fitting a given orbit, and it is an extension of that for ellipse fitting in the Kepler problem. This procedure begins with using the inner and outer radii to determine the eccentricity $e$, and semi-latus rectum, $p$. The eccentricity expansion of the p-ellipse solution gives an approximation to the ratio of the precession frequency to the orbital frequency ($m$) in terms of the potential index ($\delta$), see equation \eqref{eq4}. However, for orbits with large eccentricities $m$ is also a function of $e$. The dependences of $m$ on $\delta$ and $e$ are separable, and the latter is generally small and given by the correction factor of equations \eqref{eq14} and \eqref{eq15}. These corrections are more accurate than those derived from second-order (equation \eqref{eq8}), or third-order (see the appendix) perturbation expansions. Indeed, they reveal a dependence of the apsidal precession of the form $log(1-e)$ at high eccentricity not previously known.

A further improvement to the approximation can be obtained by adding a (first) harmonic term to the solution, as in equations \eqref{eq5} and \eqref{eq10} in the perturbation analysis, or better, equation \eqref{eq16} in the multi-point orbit fitting case.

Even the multi-point fitting to a p-ellipse plus harmonic solution given in Section 3.4 does not yield accurate fits for nearly radial orbits. Moreover, one possible fix, adding more harmonics tends to generate unstable solutions. Instead it turns out that moderate corrections to the parameter values of the fundamental plus single harmonic solutions do yield remarkably accurate fits to nearly radial orbits in flat or declining rotation curve potentials. Specifically, the correction to the precession frequency parameter $m$ discussed above remains generally valid, but needs small refinements for very eccentric orbits. However, the values of the coefficients $e_1$ and $e_2$ given by equations \eqref{eq18} and \eqref{eq19} do require further correction. In the case of the flat rotation curve, logarithmic potential this correction is given by equations \eqref{eq20} and \eqref{eq21}. Remarkably, the latter equation is nearly quadratic.  The correction will be smaller for declining rotation curve potential. The case of nearly radial orbits in rising rotation curves is more difficult and requires further research.

Nonetheless, with these corrections excellent p-ellipse fits to orbits with large radial ranges in a span of potentials are obtained. Given these properties the solutions should be useful in many applications. The first will be conversion between solution parameters ($p, e_1, e_2$) and orbital properties like maximum and minimum radii, or specific angular momentum and energy. This can aid in the interpretation of ensembles of orbits specified by any set of these quantities, just as in the Kepler problem. A second type of application is to produce analytic models of galaxy discs following a disturbance, such as an impulsive interaction as in \citet{ss12}. The accuracy of the approximations allows this even in cases when some orbits are strongly perturbed, like those that make up tidal tails. A third application is to provide an ensemble of orbits with well understood properties for producing model galaxies. These last two applications are somewhat limited by the fact that the results above apply only to single-component, spherically symmetric potentials. 

A fourth application is using the approximations to provide a means of discovering resonant orbits. As noted in Paper I, for low eccentricity orbits, equation \eqref{eq4} already suggests the existence of `resonant potentials' when the right hand side of the $m_1$ equation yields a rational number, and thus, closed orbits at low eccentricity. Additionally, as noted in connection with Fig. 6b, the eccentricity dependences of equation \eqref{eq14} guarantee that some nearly radial orbits will have rational values of $m_{corr}$ in any of the potentials considered here. For example, in the logarithmic potential $m_{corr} = 3/2$, when $e \simeq 0.918$, or $m_{corr} = 5/3$, when $e \simeq 0.999$. On the other hand, this eccentricity dependence suggests that most high eccentricity orbits are not resonant (closed) in the `resonant potentials.'

These results clarify the Bertrand Theorem, which states that only with the inverse square and Hooke's Law forces are all bound orbits closed. The fact that other potentials also have a number of closed orbits was also clear in Paper I, and another method for finding them at discrete values of energy and angular momentum was given in \citet{re10}. Equations \eqref{eq14} and \eqref{eq15} suggest that $m_{corr}$ ranges from $m_o$ to infinity, encompassing an infinity of rationals, and thus, closed orbits, though most are very nearly radial. On the other hand, in the resonant potentials (e.g., $\delta = 1/9, m_o = 4/3$), the formulae indicate an infinity of non-closed orbits at large eccentricities, so Bertrand is strictly correct, despite the fact that most orbits (in the sense of a larger infinite set) are closed in such potentials.

The value $m_{corr} = 2$ is only reached at very high values of the eccentricity for flat or falling rotation curve potentials. However, for rising rotation curve potentials, this value is attained at more modest eccentricities, decreasing as $\delta$ goes from 0 to -1. Such orbits are closed, symmetrical ovals that could form the backbone of galactic bars, and their existence opens the door to analytic, kinematic bar models in any of these potentials. This topic will be explored further in a future paper. 

Finally, the success of p-ellipse plus harmonic approximations provides some general lessons about the nature of orbits in the power-law potentials, and of how conservation law constraints mold orbits in these cases. A primary lesson is that while the ellipse solution to the Kepler problem sings in a pure tone, the orbital frequency, this is an anomaly. Generically, orbits in these potentials are primarily two-tone duets (orbital and precession frequency are not equal). A new lesson here is that the first harmonic can also be significant, if not as loud as the others. Other harmonics may be present, but the fact that the orbits can be fit so well without them, and that their presence can cause instability, suggests that in most cases the background chorus is very quiet. This may not be the case in multi-component, non-spherical potentials. The results above suggest some techniques for investigating this.

\section*{Acknowledgments}

I am grateful to B. J. Smith, S. R. Valluri and D. Lynden-Bell for their continuing interest in this topic, and the impetus provided by their communications and publications. I acknowledge use of NASA's Astrophysics Data System.

\bibliographystyle{mn2e}

\section{Appendix: Expressions in the third-order approximation}

In this appendix we give the coefficient equations, and the parameter expressions derived from them, in the third-order approximation. Parameter terms in the coefficients in equation \eqref{eq11} occur repeatedly in this analysis, so it is convenient to define notations for them,

\begin{multline}
\label{eq22}
d_a = \frac{1}{2} + \delta,\ \ 
d_b = \left( \frac{1}{2} + \delta \right) \left( 2h_2 -\frac{1}{2} + \delta \right),\\
d_c = \left( \frac{1}{2} + \delta \right) \left( \frac{-3}{2} + \delta \right)
\left( 2h_2 -\frac{1}{2} + \delta \right) \left(6h_3 + 4\delta h_2 \right).
\end{multline}

\noindent Then the coefficient equations are,

\begin{equation}
\label{eq23}
1 + d_b e^2 m^2 = \frac{c_\delta}{p_2^{2\left(\delta - 1 \right)}} = c',
\end{equation}

\begin{equation}
\label{eq24}
\left(1 - \frac{d_c}{d_a} e^2 \right) m^2 =
1 + \left(1 - 2\delta \right) c',
\end{equation}

\begin{equation}
\label{eq25}
m^2 = \frac{1}{4} + \frac{1}{4} \left(1 - 2\delta \right) 
\left[1 + 2\left( \delta -1 \right) \frac{d_a}{d_b} \right] c',
\end{equation}

\noindent and,

\begin{multline}
\label{eq26}
m^2 = \frac{1}{9} + \frac{1}{9} \left(1 - 2\delta \right)\\ 
\left[1 + 4\left( \delta -1 \right) \frac{d_a d_b}{d_c} 
2\left( \delta -1 \right) \left(2 \delta -3 \right) \frac{d_a^3}{d_c} \right] c'.
\end{multline}

\noindent These four equations are then solved for the variables: $m, c'$ (or $p$), $h_2$, and $h_3$ in terms of $e$ and $\delta$. The first result is a high order polynomial equation for $c'$,

\begin{multline}
\label{eq27}
\left( 2\delta - 1 \right) c' = 1+ m^2  \\
 \times \left\{ 2c' \left( 2\delta - 1 \right)
\left( \delta - 1 \right) \left[ 2D_1
+  \left( 2\delta - 3 \right) 
\left( \frac{1}{2} + \delta \right) \right] D_2
- 1 \right\}
\end{multline}

\noindent with,

\begin{eqnarray}
\label{eq28}
D_1 = \frac{4 \left( 2\delta - 1 \right) \left( \delta - 1 \right) c'}
{3 - 12m^2 + 2 \left( 1 - 2\delta \right) c'} ,\\
D_2 = \frac{\left(\frac{1}{2} + \delta \right) e^2}
{-9m^2 + 1 + \left( 1 - 2\delta \right) c'} ,\\
m^2 = \frac{1}{4}
\left[ \frac{ \left( c' - 1 \right) \left[ 3 + 2 \left(1 - 2\delta \right) c' \right]}
{3 \left( c' - 1 \right) + \left( 2\delta - 1 \right) \left( \delta - 1 \right)
\left( \frac{1}{2} + \delta \right) e^2 c'} \right].
\end{eqnarray}

\noindent These two equations yield $c'$ and $m_2$, and then the following equations for $d_b$ and $d_c$ can be obtained from the coefficient equations above, i.e.,

\begin{equation}
\label{eq31}
d_b = \frac{4 \left( 2\delta - 1 \right) \left( \delta - 1 \right) d_a c'}
{3 - 6m^2 +2 \left(1 - 2\delta \right) c'},
\end{equation}

\noindent and,

\begin{equation}
\label{eq32}
d_c = \frac{2 \left( 2\delta - 1 \right) \left( \delta - 1 \right)
\left[ 2d_b + \left( 2\delta - 3 \right) d_a^2 \right] d_a c'}
{1 - 9m^2 +2 \left(1 - 2\delta \right) c'},
\end{equation}

\noindent  Then, from the definitions in equation (26) we can derive $h_2$ and $h_3$, the third order coefficients.

\bsp
\label{lastpage}
\end{document}